# Realization of AlSb in the double layer honeycomb structure: a robust new class of two-dimensional material

Le Qin[1,&], Zhi-Hao Zhang[1,&], Zeyu Jiang[2,&], Kai Fan[1], Wen-Hao Zhang[1], Qiao-Yin Tang[1], Hui-Nan Xia[1], Fanqi Meng[3], Qinghua Zhang[3], Lin Gu[3,4,5], Damien West[2,#], Shengbai Zhang[2], Ying-Shuang Fu[1,*]

1. School of Physics and Wuhan National High Magnetic Field Center, Huazhong University of Science and Technology, Wuhan 430074, China

2. Department of Physics, Applied Physics & Astronomy, Rensselaer Polytechnic Institute, Troy, New York 12180, USA

3. Beijing National Laboratory for Condensed Matter Physics, Institute of Physics, Chinese Academy of Sciences, Beijing 100190, P. R. China.

4. School of Physical Sciences, University of Chinese Academy of Sciences, Beijing 100049, China.

5. Songshan Lake Materials Laboratory, Dongguan, Guangdong 523808, China.

[&] These authors contribute equally to this work.

Emails: [#]damienwest@gmail.com, *yfu@hust.edu.cn

**Exploring new two-dimensional (2D) van der Waals (vdW) systems is at the forefront of materials physics. Here, through molecular beam epitaxy on graphene-covered SiC(0001), we report successful growth of AlSb in the double-layer honeycomb (DLHC) structure, a 2D vdW material which has no direct analogue to its 3D bulk and is predicted kinetically stable when freestanding. The structural morphology and electronic structure of the experimental 2D AlSb are characterized with spectroscopic imaging scanning tunneling microscopy and cross-sectional imaging scanning transmission electron microscopy, which compare well to the proposed DLHC**

**structure. The 2D AlSb exhibits a bandgap of 0.93 eV versus the predicted 1.06 eV, which is substantially smaller than the 1.6 eV of bulk. We also attempt the less-stable InSb DLHC structure; however, it grows into bulk islands instead. The successful growth of a DLHC material here opens the door for the realization of a large family of novel 2D DLHC traditional semiconductors with unique excitonic, topological, and electronic properties.**

Two-dimensional (2D) vdW materials not only host a number of exotic physical phenomena, but provide promising candidates for applications in next generation electronics [1]. Pioneered by graphene, whose low energy electronic states support Dirac fermions, recent progress has realized numerous categories of vdW single layers [2,3] which have demonstrated unique phenomena that are unprecedent in their bulk counterpart [4-8]. These are exemplified in spin and valley polarization in monolayer transition metal dichalcogenides [8], intrinsic 2D ferromagnetism in single layer $CrI_3$ [9], as well as topological phase in single layer 1T'-transition metal dichalcogenides [10] and elemental monolayer crystal of stanene [11,12] and bismuthene [13]. The exploration of new 2D materials has been persistently providing fertile ground for the uncovering of novel quantum phenomena.

Monolayer vdW materials are routinely obtained with top-town exfoliation techniques [14], wherein preexisting weakly bound 2D sheets are mechanically removed from the bulk 3D network. While the exfoliated monolayers are of comparably high quality, the process is of low efficiency and only applicable to parent crystals with weak vdW interlayer interactions. On the other hand, bottom-up thin film growth methods, such as chemical vapor deposition or molecular beam epitaxy (MBE), are capable of synthesizing large-scale films with controlled thickness [15]. More importantly here, however, the grown films may exhibit crystal structures that are completely different than their bulk counterparts. For instance, 1T-NbSe$_2$ and 1T'-WSe$_2$ only exist in thin films [16,17], whose unit layer in the bulk is the 1H structure. Such bottom-up growth techniques open the possibility of growing novel 2D phases of materials whose bulk 3D crystal structures are strongly covalent, for example technologically mature zinc-blende (ZB) or wurtzite semiconductors.

The ability to grow 2D vdW phases of materials which are typically associated with covalent 3D bulk would greatly increase the available candidates of 2D materials for exploring new physics and functionalities for device applications. Recently, some of the authors have theoretically proposed that a large family of traditional binary semiconductors are stable in the 2D double layer honeycomb (DLHC) structure [18]. Namely, traditional III-V, II-VI, and I-VII semiconductors, with wurtzite or ZB structure in their 3D bulk form, can be kinetically stable as 2D materials at the ultrathin limit. These newly proposed 2D materials are intrinsically stable, in distinct contrast to those 'extrinsic' ones such as stanene and bismuthene, which are in essence an extension of the surface as their interaction with the substrate is required to prevent buckling [19]. Furthermore, this new class of 2D material is not only large but expected to exhibit rich phenomena, including a novel type of topological insulator based on vdW interlayer interactions [18] and excitonic insulators driven by reduced screening [20].

In this work, we attempt to grow the traditional semiconductors, AlSb and InSb, in the theoretically proposed 2D DLHC phase through MBE. The substrate of choice is graphene covered single crystal SiC(0001), in which the weak vdW interaction of graphene with the grown materials makes it feasible to examine the intrinsic nature of the resulting 2D grown films. Herein, we determine the set of growth parameters which enable the growth of vdW layered AlSb films down to the monolayer limit. Additionally, the characterized structure and electronic properties of the resulting monolayer AlSb via spectroscopic imaging scanning tunneling microscopy (STM) and cross-sectional imaging scanning transmission electron microscopy (STEM) are completely consistent with the proposed DLHC structure. This is in contrast, however, to our attempts to grow monolayer InSb. While we fully explored the available growth parameters, the growth of vdW layered InSb seems unattainable, and instead InSb is found to robustly grow into 3D islands which preserve their bulk 3D structure.

**Theory of the DLHC structure.**

Traditional I-VII, II-VI, and III-V bulk semiconductors overwhelmingly exist in either the ZB or wurtzite configurations and have a characteristic strong tetrahedral covalent bonding network. However, as semiconductor films become thinner and thinner, approaching the nanoscale, the surface energy associated with these bulk semiconductors becomes an increasingly large fraction of the total system energy. As such, in the 2D limit, novel structures and phases with lower surface energies may naturally become more energetically favorable. Indeed, we have recently conducted a survey of 28 traditional semiconductors using first-principles density functional theory (DFT) and found that in the ultra-thin limit the vast majority reduce their surface energy by forming a novel vdW layered structure, the DLHC structure [18].

DFT calculation [see Methods] of the energetics of AlSb and InSb show that both materials have a window of thickness in which the DLHC is more stable than their bulk ZB structures. The relaxed structures corresponding to ZB and DLHC for a film of AlSb (4 atomic bilayers thick) are shown in Figs. 1 (a) and (b), respectively. The DLHC structure consists of pairs of buckled honeycomb sheets which are stabilized through dative bonding to form a quadruple layer. These resulting quadruple layers are then only very weakly bound through vdW interaction. Unlike traditional semiconductors in the honeycomb structure, calculation of the phonon dispersion [18] for monolayer DLHC AlSb and InSb, shown in Fig. S1, have an absence of soft phonon modes, indicating that they are kinetically stable.

While bulk DLHC is higher in energy than bulk ZB, the formation of vdW layers all but eliminates the surface energy associated with being an ultra-thin film. As such, the formation energy of DLHC varies only weakly with number of layers, yielding a crossover in the stability with DLHC being more stable for very thin films and ZB being more stable for thicker films. The formation energies of DLHC and truncated ZB for AlSb and InSb relative to bulk ZB are shown in Figs. 1 (c) and (d), respectively. Here it can be seen that for less than 10 AlSb bilayers (film thickness of approximately 3 nm) the DLHC structure of AlSb is more stable than the correspond bulk structure. While InSb is also stable in the DLHC form, it is substantially less so, with a single DLHC quadruple layer being 260 meV more stable per formula unit than the corresponding bulk structure compared with the 510 meV of AlSb. The diminished stability of InSb is also seen in the crossover, which occurs at thickness of less than 2 nm with 6 bilayers of InSb in the ZB form being more stable than 3 quadruple layers of DLHC InSb.

**Growth of AlSb islands and films.**

The growth of AlSb has been carried out on a Si substrate in previous studies [21,22], in which it was found that only 3D islands with bulk structures are obtained. Here, we revisit its growth on a graphene substrate. Different to Si, the graphene is inert and only weakly interacts with the grown films, rendering it not only devoid of reaction with AlSb but also having minimal strain accumulation due to lattice mismatch at the interface. Our growth of AlSb is performed by evaporating Al and Sb with an approximate flux ratio of 1:5. The substrate temperature is kept relatively high to facilitate the film formation and desorption of excess Sb. This strategy of sample growth, which is inherited from the growth of traditional GaAs [23], has been successfully applied in numerous binary vdW compounds [24-26]. Under the condition of excess Sb, the growth behavior of AlSb is mainly determined by the substrate temperature ($T_s$) and the beam flux of Al. Those two parameters are correlated. Namely, higher Al flux corresponds to decreasing $T_s$. We therefore vary $T_s$ with a fixed Al flux. Figure 2(a) shows the sample morphology grown with a $T_s$ of 300 °C. AlSb islands with 3D features can be seen, with heights of 4 nm - 8 nm and lateral sizes of 15 nm - 30 nm. The island surfaces have a roughness of 0.5 nm, apparently not of atomically flat.

As $T_s$ is lowered to 250°C, large 2D AlSb films, coexisting with 3D islands, appear with extensions of ~100 nm [Fig. 2(b)]. The 2D film is atomically flat and has a height of approximately 1 nm. The trace clusters, randomly distributed on the film surface, can be reduced with extended post-annealing after the sample growth, which facilitates their crystallization, but have not been fully eliminated. The appearance of those clusters maybe related to insufficient diffusion on AlSb compared to that on graphene, however, we note that coexisting 3D islands share similar sizes and heights as those of Fig. 2(a). With further decreasing $T_s$ to 170 °C, the sample surface is dominated with 2D AlSb films of uniform height [Fig. 2(c)]. Only a small fraction of residual islands of ~ 1.4 nm in height remain. When $T_s$ is below 170°C, the sample surface is composed of 3D islands again without noticeable regularity [Fig. 2(d)]. It is possible that the substrate temperature is not high enough to allow adequate reaction between Al and Sb. Thus, we find 170°C to be the optimal substrate temperature for the growth of 2D AlSb films.

Having identified the optimal $T_s$, we further investigate the coverage dependence of the growth behavior of the AlSb films. This is not only required for characterization of its physical properties with ensemble averaged techniques but is also relevant to building functional electronic or optical devices. Fig. 3(a) shows the morphology of the as-grown AlSb films after extended post annealing, whose nominal coverage is estimated as 0.63 monolayer (ML). Note that one ML is refers to one quadruple layer of the DLHC structure of AlSb. While the coverage of the 2D films becomes larger compared to that shown in Fig. 2(c), the traces of protruding AlSb clusters/islands residing on top of the films also become more abundant. This results in the exposed 2D film surfaces constituting approximately 56% of the field of view (FOV). With further increasing the nominal coverage to 0.95 ML, the 2D films constitute 72% of the FOV [Fig. 3(b)], meanwhile the protruding islands increase to 28% of the FOV, essentially leaving ~44% of the FOV exposed with the film surfaces. As shown in Fig. 3(d), the measured heights of the islands on top of the 2D films are not uniform, ranging from ~1.15 nm to 4.6 nm above the film surfaces. With the nominal coverage further increasing to 1.26 ML, the exposed 2D films reduce to ~38% of the FOV, due to the increasing density of the islands [Fig. 3(c)]. This demonstrates more layers of AlSb starts to grow before the completion of the 2D films. The optimum coverage for exposing the film surfaces is around 0.6 ML.

**Structure and morphology of single layer AlSb.**

With the achieved growth of 2D AlSb films, we characterize their structural morphology and electronic states in combination with first-principles calculations. Figure 4(a) shows the theoretically proposed DLHC structure of monolayer vdW AlSb. It is transformed from a truncated bilayer 3D bulk AlSb in the (111) direction by lateral shifting of the Al layer beneath the top Sb layer. Alternately, the structure may be viewed as two buckled honeycombs, with Sb protruding from both sides, which are stabilized due to their strong mutual interaction. Therefore, the top Sb layer is expected to dominate the tunneling current of STM imaging, giving rise to a triangular lattice with a constant of 0.429 nm. The height of the DLHC structure is 0.396 nm. Accounting for the vdW interlayer distance to the substrate, the relative height between the top Sb layer and the graphene surface is calculated as 0.767 nm.

Fig. 4(b) shows a selected 2D AlSb film imaged at 3 V. Its height is measured as 0.844 nm from a traversing line profile across the film edge to the graphene substrate [Fig. 4(c)]. This apparent height is in line with the theoretical height of a DLHC AlSb on graphene. More importantly, zoom-in STM image to the AlSb film resolves its atomic resolution, which indeed exhibits a triangular lattice [Fig. 4(d)]. From its fast Fourier transformation (FFT) [Fig. 4(e)], the lattice constant is determined as 0.42 nm, in reasonable agreement with the theoretical value of 0.429 nm. This is in distinct contrast to its bulk form, which exhibits surface reconstructions [28]. These observations demonstrate the 2D AlSb film clearly possess the expected structure of DLHC. Notably, the FFT image also resolves a superstructure with a period of 1.53 nm (marked with yellow circle). It originates from a $\sqrt{13} \times \sqrt{13}$ moiré pattern between AlSb DLHC and graphene interface, which has an expected period of 1.547 nm and agrees well with the experiment (Fig. S2). The emergence of the moiré pattern gives another support of the DLHC structure of AlSb.

The apparent height of the AlSb film determined from STM imaging is influenced by the electronic structure. To disentangle such influence, we characterized the structure of the AlSb film with cross-sectional imaging STEM. Fig. 4(f) shows a STEM image of the AlSb film capped with a protection layer of ~10 nm Sb. The STEM image displays the interfacial atomic structures, including the SiC substrate, the graphene layer, the AlSb film, and the Sb capping layer. The Sb atoms have much stronger contrast than those of the Al atoms, because of their heavier mass. The atomic resolution of the AlSb films is not as evident as that of the SiC substrate, due to the short-projected distance between Al and Sb atoms. Nevertheless, the STEM image of the AlSb film evidently display two atomic layers, which match the expected DLHC structure reasonably well. The measured height of the AlSb film relative to the graphene surface is about 0.76 nm, in excellent agreement with the calculation and also conforming to the STM measurement.

**Electronic structure of single layer AlSb.**

Next, we study the electronic structure of the 2D AlSb film. Fig. 5(a) shows tunneling conductance spectrum, corresponding to density of states of the film. It is an averaged spectrum from a series of spectra [Fig. S3(b)] taken along a black line shown in Fig. S3(a). It features an evident insulating gap of 0.93 eV. The conduction and valence band edges are located at 0.47 eV and -0.46 eV, respectively. The conductance intensity monotonically increases with decreasing energy below the valence band edge. For the conductance bands, there exhibit two peaks at 0.62 eV and 0.74 eV, superimposed on the continuously increasing conductance above the conduction band edge. The 2D conductance plot in Fig. S3 indicates the electronic states of the AlSb film is approximately uniform in real space. This implies the high crystalline quality of the grown film.

To obtain the momentum resolved band structure of the 2D AlSb, we further perform spectroscopic imaging STM to an island [Fig. 5(b), Fig. S4(a)]. The step edges of the island act as scattering centers for the electronic states of AlSb. The quasi-particle scattering generates prominent standing wave patterns, whose wavelengths depend on the imaging energy, as is exemplified in the selected conductance images shown in Fig. 5(c). By performing FFT to the conductance images, scattering wave vectors shown in $q$ space are obtained [29]. The FFT signal can be further enhanced by averaging the raw FFT pattern through the threefold symmetry of the in-plane crystal structure of AlSb. Since the scattering wave vector $q$ connects momentum wave vector $k$ on the constant energy contour, the FFT pattern reflects the band structure of the material. Further, the energy dispersion of the band structure can be extracted from the evolution of FFT patterns with energy.

As is shown in Fig. 5(d) and Fig. S4(b), the FFT images, corresponding to those of Fig. 5(c), exhibit distinct patterns with six-fold symmetry, which disperse markedly with energy. The energy dispersions along Γ-K and Γ-M directions are extracted and plotted in Figs. 5(e) and (f), respectively. They both feature a band gap whose gap size is consistent with the point spectrum. Evidently, the dispersions of the conduction bands along the both directions are much narrower in $q$ space compared to that of the valence bands. Furthermore, the dispersion along Γ-M direction is more extended than that along Γ-K, which is well captured by our first principles calculations. The calculated density of states shows a gap of 1.06 eV [Fig. 5(g)], whose intensity increases with energy above (below) the conduction (valence) band edge. This is close to the measured gap of 0.93 eV, and consistent with the general spectroscopic trend of the conductance spectra. The calculated bands [Fig. 5(h)] also show conduction bands with a smaller effective mass of 0.10 $m_e$ compared to that of the valence bands of 0.17 $m_e$, where $m_e$ is the mass of the free electron. In comparison with the calculated bands, the experimentally measured dispersions are from intra-band scattering for the conduction and valence bands, as is indicated with black arrows in Fig. 5(h). A quantitative comparison with the experiment is performed by superimposing the theoretical intra-band scattering dispersion [red curves in Figs. 5(e) and (f)], which shows quite reasonable agreement.

**Growth of 3D InSb islands.**

Theoretical study also predicts the transformation of truncated bulk InSb to the same DLHC structure as AlSb at the 2D limit. In addition, the double layer DLHC structure of InSb is expected to host a topological insulator phase with a gap of 84 meV coming from the interlayer interaction induced band inversion, making it an interesting candidate for hosting possible high temperature spin Hall effect [18]. We therefore test the growth of InSb films on the same graphene covered SiC(0001) substrate by similar coevaporation of In and Sb with a flux ratio of ~1:5. The substrate temperature has been varied from 170 °C to 270 °C. However, the synthesized InSb robustly grows into 3D islands. Decreasing the substrate temperature merely results in the reduction of thickness and sizes of the islands (Fig. S5). The thinnest InSb island grown at 170 °C is 3.6 nm, which from Fig. 1 (d) is more than twice as thick as where the DLHC structure is predicted to be stable. Moreover, at this temperature the island surface exhibits roughening, a signature of deteriorated crystallinity. The 3D islands grown at higher temperatures, on the other hand, are all of high quality as described below.

The sample morphology of the InSb islands grown at 270 °C are exemplified in Fig. 6(a). The islands possess hexagonal or triangle shapes with straight edges. Their lateral sizes range from 170 nm to 200 nm, and their thicknesses vary from 6.5 nm to 17.5 nm. The surface of the InSb islands are atomically flat. Atomic resolution imaging of the InSb islands, display a clear 2×2 reconstruction [Fig. 6(b)]. This surface reconstruction has been observed on bulk InSb(111) surface with Sb termination [30]. The $2 \times 2$ reconstruction comes from the trimerization of the top layer Sb atoms, whose crystal structure is shown in Fig. 6(c). This agrees with our growth condition of excess Sb, and unambiguously substantiates our grown islands are of the bulk sphalerite structure. The calculated formation energy of InSb DLHC is 0.51 eV per unit formula. It is indeed considerably larger than that of AlSb, which is 0.41 eV per unit formula. Moreover, the stability of the InSb DLHC layers is even further reduced when considering the competing 2×2 reconstruction (see Fig. S6). Therefore, the unattained growth of 2D InSb films can also be understood.

In summary, we have provided the first experimental realization of a material, namely the traditional III-V semiconductor AlSb, in the theoretically predicted 2D DLHC structure. This was accomplished though MBE growth on a graphene covered SiC(0001) substrate. The atomic crystal structure of the resulting AlSb monolayer, characterized with STM and STEM, and the electronic band structure, imaged with STM based quasi-particle scattering, are in good agreement with that from first principles calculations of the DLHC phase. While additional attempts were made with InSb, our current growth approach instead led to the growth of 3D islands maintaining their bulk structure, whose surfaces express the typical InSb(111)-2×2 reconstruction. The attained growth of 2D AlSb is the first experimental verification of the large class of predicted DLHC materials which are derived from traditional binary semiconductors. The DLHC structure of AlSb has quite different properties than bulk and is expected to have improved carrier mobility and enhanced excitonic binding [31]. This work paves the way toward the realization of other members of this large family of 2D materials and further investigations into their properties and exotic electronic phenomena.

## MATERIALS AND METHODS

**Sample preparation.**

The SiC substrate (MTI corporation) was firstly degassed at 900 K for a minimum of 3 hours. Then, it is flashed to 1220 K for 2 min while facing a Si source heated at 1470 K. After five flashing cycles, the Si source is turned off. And the SiC is flashed to 1670 K for 5 min to desorb Si atoms and form a graphene layer [32].

The AlSb films are grown by coevaporation of high purity Al (purity 99.999%) and Sb (purity 99.95%) from a tungsten wire basket coated with alumina and a Knudsen-cell evaporator, respectively. The growth rate of AlSb is ~ 0.028 monolayer per minute. The temperature of the graphene-covered SiC substrate is controlled with current heating and monitored with an infrared thermometer.

The InSb islands are grown similarly as AlSb by coevaporation of high purity In (purity 99.999%) and Sb (purity 99.95%) from a tantalum boat and a Knudsen-cell evaporator, respectively. The growth rate of InSb is ~0.043 monolayer per minute.

**STM measurement.**

The experiments were performed with a home-built room temperature STM and a custom-made cryogenic Unisoku STM at 5 K [33]. Both systems are equipped with MBE. An electro-chemically etched W wire was used as the STM tip. Prior to measurements, the tip was characterized on Ag(111) multi-layer films grown on a Si(111) substrate, which has been cleaned by several flashing cycles to 1500 K. The tunneling spectra were obtained by lock-in detection of the tunneling current with a modulation voltage at 983 Hz feeding into the sample bias.

**STEM imaging.** The cross-section STEM specimens were prepared using a routine focused ion beam. The MBE-grown AlSb films were further capped with ~10 nm Sb layers as protection layers for ex-situ STEM analysis. The atomic structures of the AlSb films was characterized using an ARM－200CF (JEOL, Tokyo, Japan) transmission electron microscope operated at 200 kV and equipped with double spherical aberration (Cs) correctors. ADF images were acquired at acceptance angles of 40-160 mrad. All imaging procedures were performed at room temperature.

**First principles calculation.**

The crystalline structures were obtained in the framework of DFT with Perdew-Burke-Ernzerhof (PBE) exchange-correlation functional as implemented in VASP [34]. Lattice and atomic degrees of freedom were fully relaxed with the consideration of spin-orbital coupling, until the residual forces were less than 0.002 eV/Å. A DFT-D3 type vdW correction was employed to describe the interaction between AlSb film and graphene substrate [35]. A vacuum of 30 Å was used to avoid fictitious interactions from periodic boundary condition. Electronic band structure was calculated by HSE@$G_0W_0$ approach. Namely, it was obtained firstly by a hybrid functional HSE calculation [36], then further improved by a quasi-particle $G_0W_0$ correction. An energy cut-off of 600 eV and *k*-point grid of 12×12×1 were employed for all the calculations.

**Author contributions:** Y.S.F., D.W. and S.B.Z. conceived the project. L.Q., Z.H.Z. and K.F. performed the experiments with the help of W.H.Z., Q.Y.T. and H.N.X.. Z.J., and D.W. did the first principles calculations. F.Q. M, Q.H. Z and L. G performed the STEM characterization. Y.S.F., L.Q., Z.J., D.W. and S.B.Z. analyzed the data. Y.S.F., D.W. and S.B.Z. wrote the manuscript. All authors discussed the data and commented on the manuscript.

**Acknowledgement:** We thank Yuanchang Li, Junbo Han, and Tianyou Zhai for helpful discussions. Work in China is funded by the National Key Research and Development Program of China (Grant Nos. 2017YFA0403501, 2016YFA0401003, 2018YFA0307000), the National Science Foundation of China (Grant Nos. 11874161, U20A6002, 11774105, 52025025, 52072400). Work in the US was supported by the U.S. DOE Grant No. DESC0002623. The supercomputer time sponsored by NERSC under DOE Contract No. DE-AC02-05CH11231 and the CCI at RPI are also acknowledged.

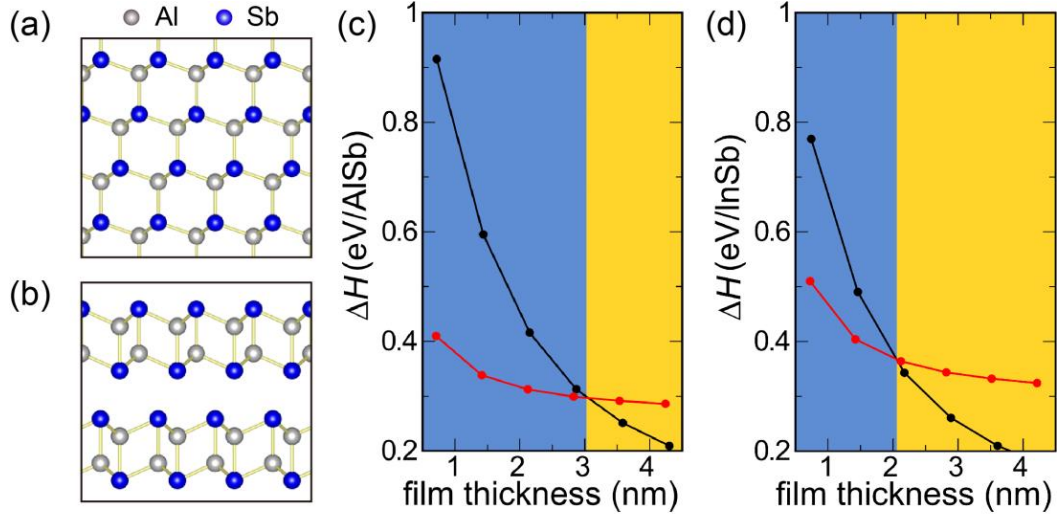

**Fig. 1. Structure and energetic stability of the predicted DLHC structure.** (a,b) The DFT relaxed atomic structure of a thin film (sectional view) containing 4 AlSb bilayers in the (a) bulk ZB structure and (b) the vdW layered DLHC structure. (c,d) The formation energy per formula unit of the ZB (shown in black) and DLHC (shown in red) structures, relative to the bulk crystal, as a function of film thickness for AlSb (c) and InSb (d). The blue region indicates the thickness in which the DLHC structure is more stable than the corresponding truncated bulk ZB structure.

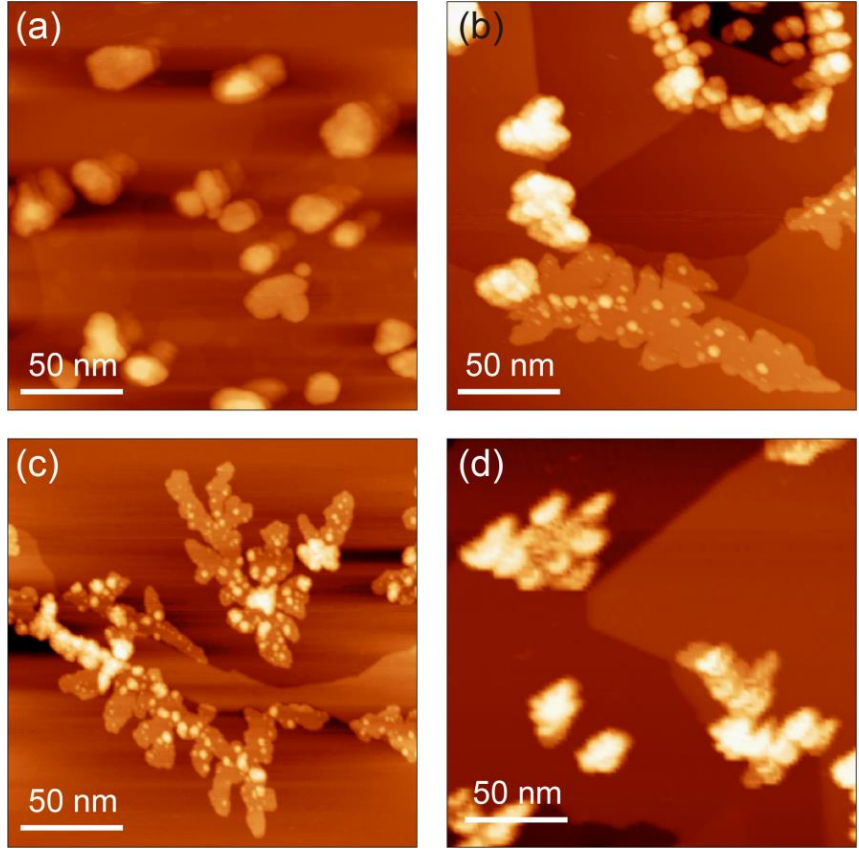

**Fig. 2. Substrate temperature dependent growth of AlSb islands and films.** (**a-d**) STM topography of the as grown AlSb islands and films with different substrate temperature $T_s$ during growth. Substrate temperature: (**a**) $T_s = 300$ °C; (**b**) $T_s = 250$ °C; (**c**) $T_s = 170$ °C; (**d**) $T_s < 170$ °C. Imaging conditions: (**a**) $V_b = 3$ V, $I_t = 10$ pA, (**b**) $V_b = 2$ V, $I_t = 5$ pA, (**c**) $V_b = 3$ V, $I_t = 5$ pA, (**d**) $V_b = 3$ V, $I_t = 5$ pA. Note that there is a double tip in the image of (b) for the high islands.

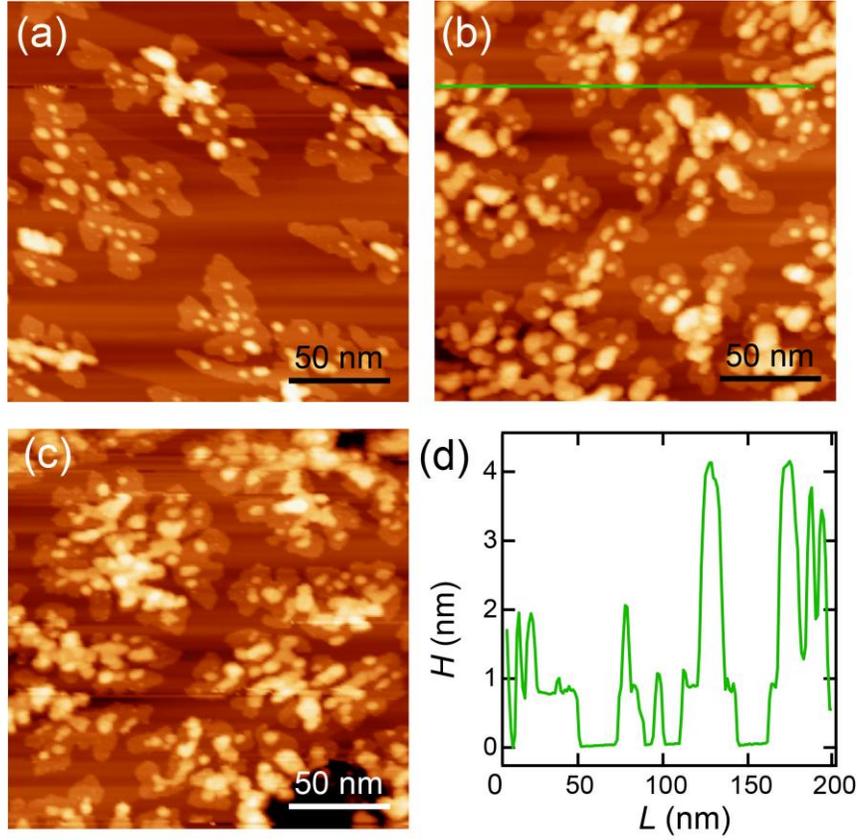

**Fig. 3. Coverage dependent growth of AlSb films.** (**a-d**) STM topography of the as grown AlSb films after post annealing with different nominal coverages $\theta$. Nominal coverage: (**a**) $\theta = 0.63$ ML; (**b**) $\theta = 0.95$ ML; (**c**) $\theta = 1.26$ ML. Imaging conditions: (**a**) $V_b = 3$ V, $I_t = 5$ pA, (**b**) $V_b = 2.5$ V, $I_t = 5$ pA, (**c**) $V_b = 2.3$ V, $I_t = 5$ pA. (**d**) Line profile extracted along the green line in (**b**).

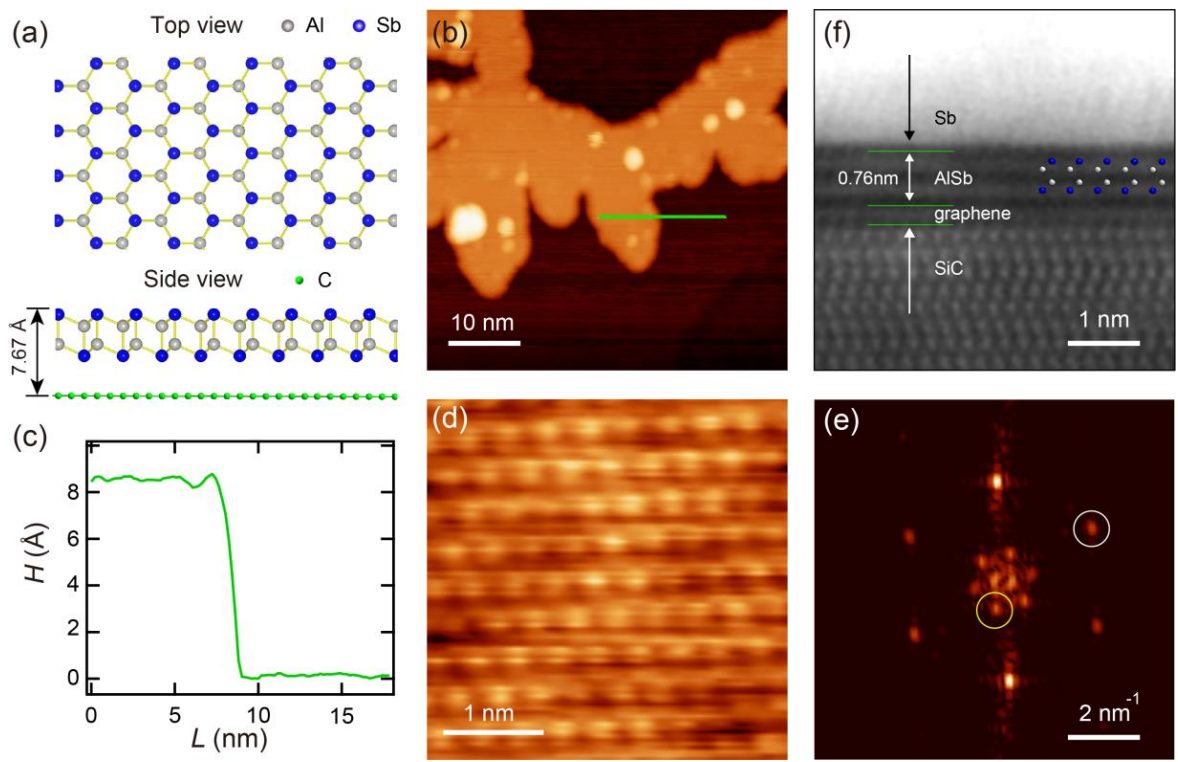

**Fig. 4. Structure and morphology of AlSb DLHC.** (a) Top and side view of the crystal structure of DLHC AlSb supported on graphene. (b) STM topography ($V_b$ = 3 V, $I_t$ = 5 pA) of a DLHC AlSb film. (c) Apparent height (*H*) of the DLHC AlSb measured along the green line in (b). (d) STM image ($V_b$ = 0.7 V, $I_t$ = 30 pA) of the atomic resolution of DLHC AlSb, (e) FFT of (d). The white circle marks the diffraction spot of the atomic resolution of the Sb lattice. The yellow circle corresponds to the Moiré superstructure between graphene and the AlSb monolayer. (f) STEM image of the cross-sectional view of the AlSb film, which is superimposed with the DLHC structural model. The Sb (Al) atoms are depicted with blue (grey) balls.

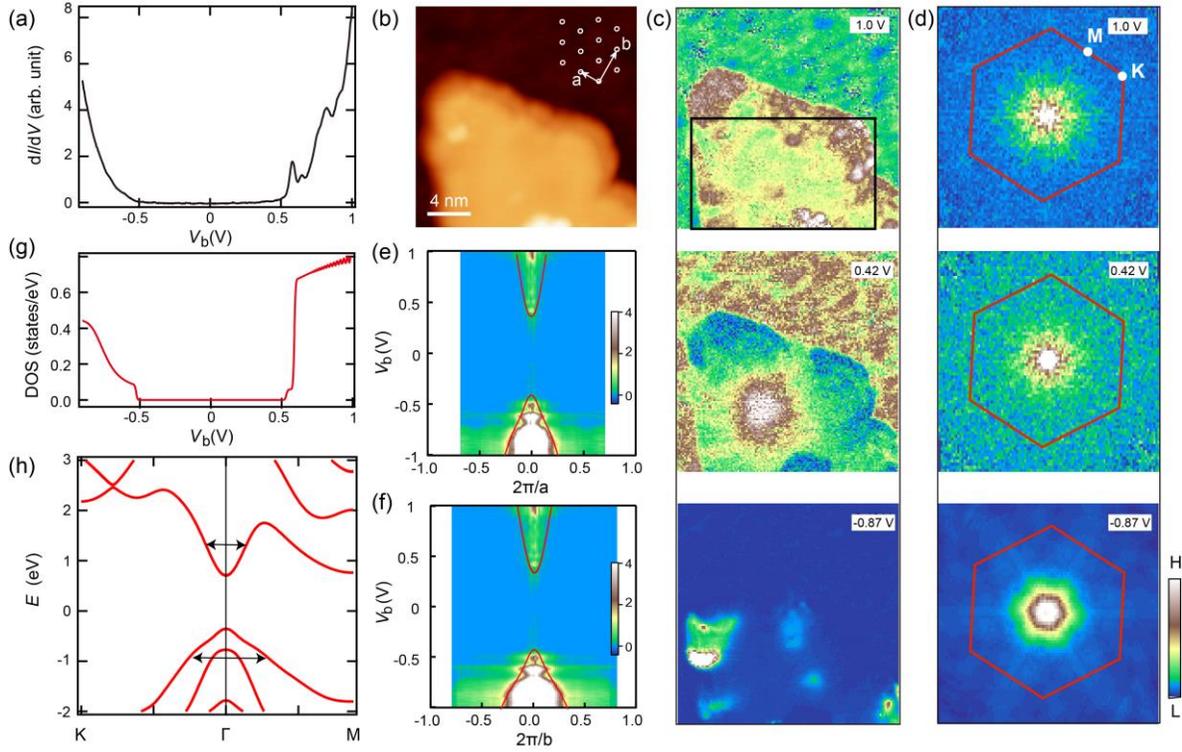

**Fig. 5. Electronic structure of DLHC AlSb.** (a) Averaged tunneling spectrum of DLHC AlSb taken along the black line in Fig. S3(a). Spectroscopic conditions: $V_b = 1$ V, $I_t = 100$ pA, $V_{mod} = 20$ mV. (b) STM topography ($V_b = 3$ V, $I_t = 5$ pA) of a DLHC AlSb island. Inset is a schematic of the top layer Sb atoms with unit vectors marked. Note that there is a double tip in the image. (c,d) Selective conductance mappings (c) of the imaged area in (a) and their corresponding FFT images (d) of the area marked with black rectangle in (c). The red hexagons mark the Brillion zone of the DLHC AlSb. Spectroscopic conditions: $V_b = 1$ V, $I_t = 100$ pA, $V_{mod} = 20$ mV. (e,f) Energy dispersion relation with the scattering wavevector extracted along the G-K (e) and G-M direction (f). The red curves depict the theoretical dispersions of the intra-band scattering marked in (h), which has been rigidly shifted in energy to compare with the experiments. (g,h) Density of states (DOS) (g) and bands (h) of DLHC AlSb, obtained by HSE@$G_0W_0$ approach (See METHODS for details). The black arrows in (h) mark the intra-band scattering channels of conduction and valence bands.

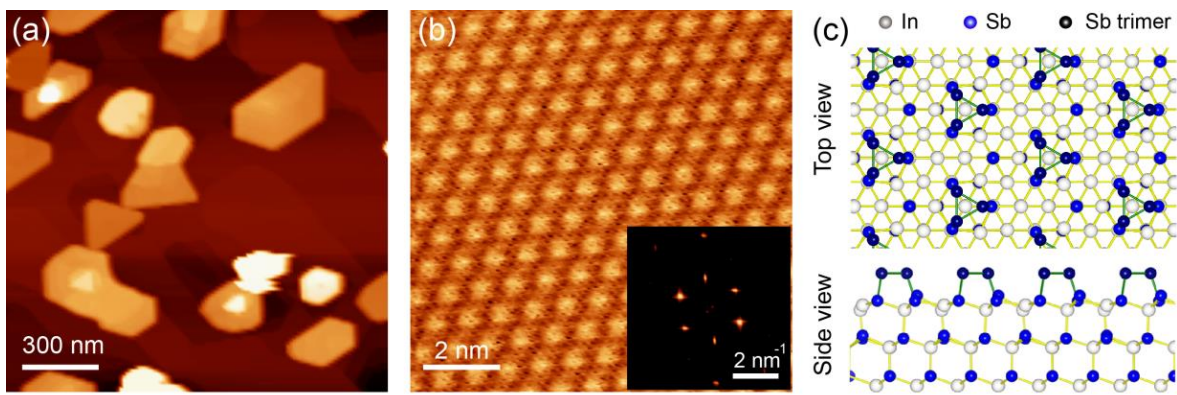

**Fig. 6. Growth of 3D InSb islands.** (a) Large scale STM image ($V_b$ = 2 V, $I_t$ = 10 pA) showing the topography of the as-grown InSb islands with regular hexagonal or triangular shape and flat surfaces. (b) Zoom-in image ($V_b$ = 2 V, $I_t$ = 5 pA) showing the atomic resolution reconstruction of InSb(111) surface. The inset is an FFT image of (b). (c) Ball and stick model showing top and side view of the crystal structure of the InSb(111)-2×2 reconstruction.